**Title:** Understanding and controlling the work function of perovskite oxides using Density Functional Theory
**Authors:** Ryan Jacobs[1], John Booske[2], and Dane Morgan[1,*]

[1]Department of Materials Science and Engineering, University of Wisconsin- Madison, Madison, Wisconsin, USA

[2]Department of Electrical and Computer Engineering, University of Wisconsin- Madison, Madison, Wisconsin, USA

*Corresponding author e-mail: ddmorgan@wisc.edu

## ABSTRACT

Perovskite oxides containing transition metals are promising materials in a wide range of electronic and electrochemical applications. However, neither their work function values nor an understanding of their work function physics have been established. Here, we predict the work function trends of a series of perovskite ($ABO_3$ formula) materials using Density Functional Theory, and show that the work functions of (001)-terminated AO- and $BO_2$-oriented surfaces can be described using concepts of electronic band filling, bond hybridization, and surface dipoles. The calculated range of AO ($BO_2$) work functions are 1.60-3.57 eV (2.99-6.87 eV). We find an approximately linear correlation ($R^2$ between 0.77-0.86, depending on surface termination) between work function and position of the oxygen $2p$ band center, which correlation enables both understanding and rapid prediction of work function trends. Furthermore, we identify $SrVO_3$ as a stable, low work function, highly conductive material. Undoped (Ba-doped) $SrVO_3$ has an intrinsically low AO-terminated work function of 1.86 eV (1.07 eV). These properties make $SrVO_3$ a promising candidate material for a new electron emission cathode for application in high power microwave devices, and as a potential electron emissive material for thermionic energy conversion technologies.

**Table of Contents Figure:**

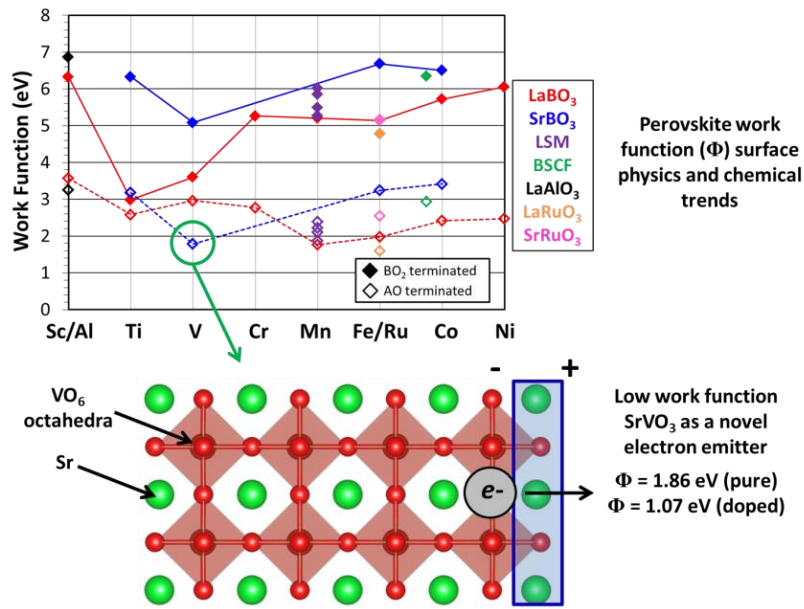

**Table of Contents Caption:** The work function values and trends of twenty technologically important perovskite oxides were investigated using Density Functional Theory. The governing work function physics were developed using concepts of bonding ionicity, hybridization, band filling, surface dipoles, and the oxygen band center as an electronic structure descriptor. SrVO$_3$ was found to be a new promising, low work function electron emission material.

# 1. Introduction

Research into the basic science and applications of perovskite oxides is extremely active and very relevant for a number of technological applications. These applications include at least oxide electronics,[1-3] catalysis and solid oxide fuel cells,[4-7] transistor dielectrics,[8, 9] field emission coatings,[10-13] magnetic tunnel junctions,[14, 15] and solid state memory.[16, 17] Perovskite materials are relevant to a wide variety of applications in part due to their stable incorporation of ~90% of the elements in the periodic table. This high degree of compositional flexibility allows for tunable properties to fit the needs of many possible applications, including the tuning of the work function. The perovskite bulk and surface structures used in this work are shown in **Figure 1,** as well as described in the Computational Methods (**Section 6**) and **Section 1** of the **Supplementary Information (SI)**. We note that we use ideal (001) surfaces without defects or atomic position reconstructions, and we discuss the impacts of this approximation in our discussion of errors between experiment and simulation in **Section 3.1** of the main paper and also in **Section 3** of the **SI**.

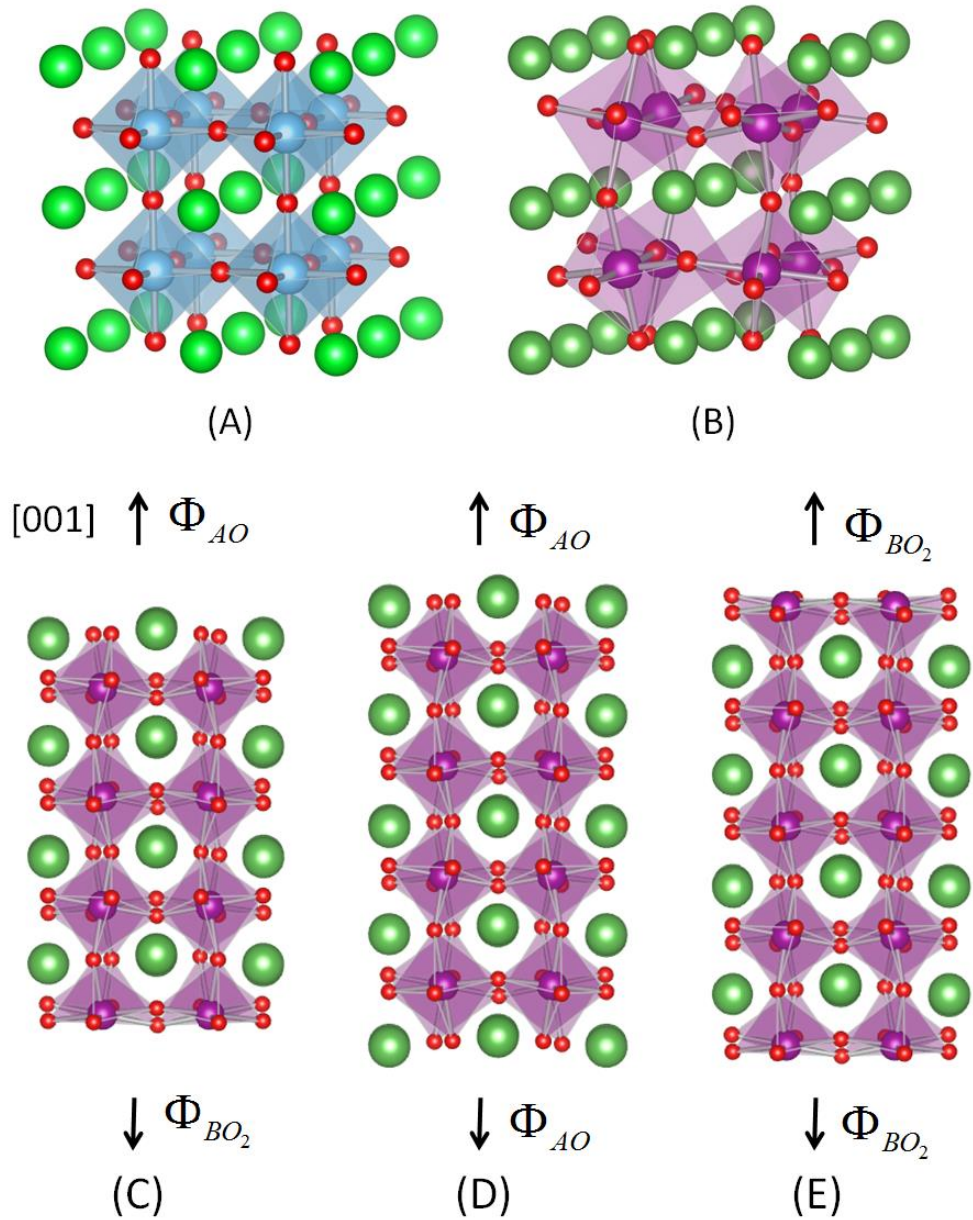

**Figure 1:** Crystal structures for (A) ideal cubic perovskite and (B) pseudocubic perovskite phases. In both (A) and (B) the green (largest, corner atoms) are the A site cations, the blue/purple atoms at the center of the octahedra are the B site cations, and the red atoms are O. These structures depict high temperature pseudocubic phases that were derived from experimental (A) $Pm\bar{3}m$ (cubic) and (B) $Pbnm$ (orthorhombic) and $R\bar{3}c$ (rhombohedral) symmetries. Structure models of $ABO_3$ surface slabs are (C): asymmetric, stoichiometric, (D): symmetric and AO terminated, nonstoichiometric, and (E) symmetric and $BO_2$ terminated, nonstoichiometric.

Knowledge of a material's work function provides an absolute electron energy level reference relative to the vacuum energy, which is important for device applications where the discontinuities of energy levels between different materials have a large effect on the device properties and performance. Absolute electron levels often play a critical role for devices with hetero-structured interfaces or active surfaces, such as solar cells, oxide electronics, electrocatalysts, and applications utilizing thermionic electron or field emission physics such as Schottky junctions, thermionic energy converters, and vacuum cathodes for high power microwave sources. Accurate work function values and an understanding of their origins and trends are critical for materials development and optimization for these classes of technologies.

In this work we provide a database and trends in work function with respect to composition changes of the A- and B-site cations of a representative set of $ABO_3$ perovskite materials, many of which may have practical value for the above applications in either pure or doped forms. Furthermore, we provide fundamental understanding of these work function trends by relating them to an electronic structure descriptor and known trends in transition metal chemistry. We use the HSE functional of Heyd, Scuseria and Ernzerhof[18] within Density Functional Theory (DFT) with Hartree-Fock exchange fractions obtained from Refs. [19, 20], which fit the exchange fractions specifically to yield correct bulk electronic properties for these materials. Use of these fitted Hartree-Fock exchange fractions ensures that the band levels and work functions are the most quantitative calculated values reported to date.

We also apply the methods and understanding gained in this work to discover a low work function perovskite material, $SrVO_3$, for efficient electron emission into vacuum (thermionic, field, or photo-emission). The main electron emission applications considered here are high power electron beam applications (such as high power microwave or millimeter-wave source technologies) and thermionic energy conversion devices. A low work function is important for electron emitters as the most facile electron removal will result in high emitted electron current densities at lower temperatures, electric fields, or light intensity in the context of thermionic, field, or photo- emission, respectively. In the rest of this paper, we specifically discuss and compare perovskite emission property predictions with conventional thermionic emitters, but the advantages are understood to apply more generally to all forms of electron emission.

Historically, thermionic electron emitters are comprised of a refractory metal such as W coated with an oxide or diffusing oxide species that lowers the work function via electrostatic surface dipoles. The coating is necessary because the refractory metals tend to have high work functions (on the order of 4.5 eV), and are therefore poor electron emitters unless a coating is included to lower their work function. Examples of thermionic emitters include impregnated W cathodes that have a low work function due to the formation of Ba-O dipoles[21] and scandate cathodes where the complex interplay between dipole formation and electron doping of Ba-O on $Sc_2O_3$ also creates a low work function.[21-23] These types of thermionic emitters are currently employed in many high power electron beam applications.[24, 25] Even thermionic energy conversion emitting layers rely on the same type of volatile surface dipole layers, such as Cs-O adsorbed on GaAs or InGaAs.[26, 27] Replacing these current emission materials which contain volatile surface species with a new material with an intrinsically low work function would simplify the architecture and increase the lifetime of electronic devices which use thermionic electron emission processes. In this work we propose that $SrVO_3$, due to its intrinsic stability, high conductivity, and low work function, is a very promising material for next generation thermionic electron emitters.

There has been some experimental and computational work related to measuring the work functions of some perovskite materials. These studies include Kelvin probe microscopy of $LaAlO_3$[2] and $A_{1-x}B_xMnO_3$ (A = La, Pr, Nd, B = Sr, Ca, Pb),[28] UPS measurements of Nb-doped $SrTiO_3$,[29] a combination of photoemission and redox potential measurements of $LaMnO_3$,[30, 31] and X-ray absorption and emission measurements on a series of transition metal-containing $LaBO_3$ (B = Cr, Mn, Fe, Co, Ni) perovskites.[32] DFT work function calculations of $BaTiO_3$,[33] $SrTiO_3$,[16] and $La_{2/3}Sr_{1/3}MnO_3$[34] have been performed by other research groups. However, none of these studies provide enough data to establish the physics and trends governing true surface work functions. Summary tables of these experimental and calculated work functions from other research groups are provided in **Section 3** of the **SI** along with a discussion of their comparison to our values. Generally, our calculated results agree with previous experimental and calculated values to the extent that comparison is possible.

# 2. Results: ABO$_3$ work function data and trends

## 2.1. ABO$_3$ calculated work functions

**Figure 2** is a plot of the calculated work functions for the AO- and BO$_2$- surface terminations versus composition of the B-site for all 20 materials considered in this work. These materials were chosen because their pure or slightly defected (doped and/or off-stoichiometric) variants have already been the subject of intense research in the areas of oxide electronics,[1-3] catalysis and solid oxide fuel cells,[4-7] transistor dielectrics,[8, 9] field emission coatings,[10-13] magnetic tunnel junctions,[14, 15] and solid state memory.[16, 17] Furthermore, these materials can exhibit the stable incorporation of many transition metals, resulting in a range of physical properties (e.g., from heavily insulating to metallic electronic conductivity), making them an ideal set of materials in which to study compositional trends in the work function. Looking from left to right, **Figure 2** shows how the AO and BO$_2$ work functions for perovskite materials change as B-site cations move across the 3$d$ series of the periodic table from Sc through Ni. **Table 1** contains the calculated work functions for the AO- and BO$_2$-terminated (001) surfaces for all ABO$_3$ materials considered in this study. The O-bond ionicities are also provided, where the ionicity is defined by the ratio of the computed atomic charge on oxygen to the value of -2 expected for a perfectly ionic system (for example, an O atomic charge of -1.5 yields a bond ionicity of -1.5/-2 = 0.75, and a perfectly ionic system would yields a bond ionicity of -2/-2 = 1). The atomic charges were calculated using Bader charge analysis of atomic charges of bulk ABO$_3$ materials.[35, 36] Finally, we have also included in **Table 1** ranges of experimentally measured work functions for select materials. Generally, chemical bonding has mixed covalent and ionic character, therefore the calculated atomic charges on oxygen from the Bader analysis will be less than (i.e., more positive than) -2. The O-bond ionicities in **Table 1** will be referenced in upcoming qualitative discussions of work function trends for these materials in **Section 3.1**. For the LaBO$_3$ series, the LaO work functions range from 1.60 eV (LaRuO$_3$) to 3.57 eV (LaScO$_3$) while the BO$_2$ work functions range from 2.99 eV (LaTiO$_3$) to 6.87 eV (LaAlO$_3$), and tend to increase in magnitude for Ti through Ni. For the SrBO$_3$ series, the SrO work functions range from roughly 1.86 eV (SrVO$_3$) to 3.42 eV (SrCoO$_3$) while the BO$_2$ work functions range from 5.09 eV (SrVO$_3$) to 6.68 eV (SrFeO$_3$).

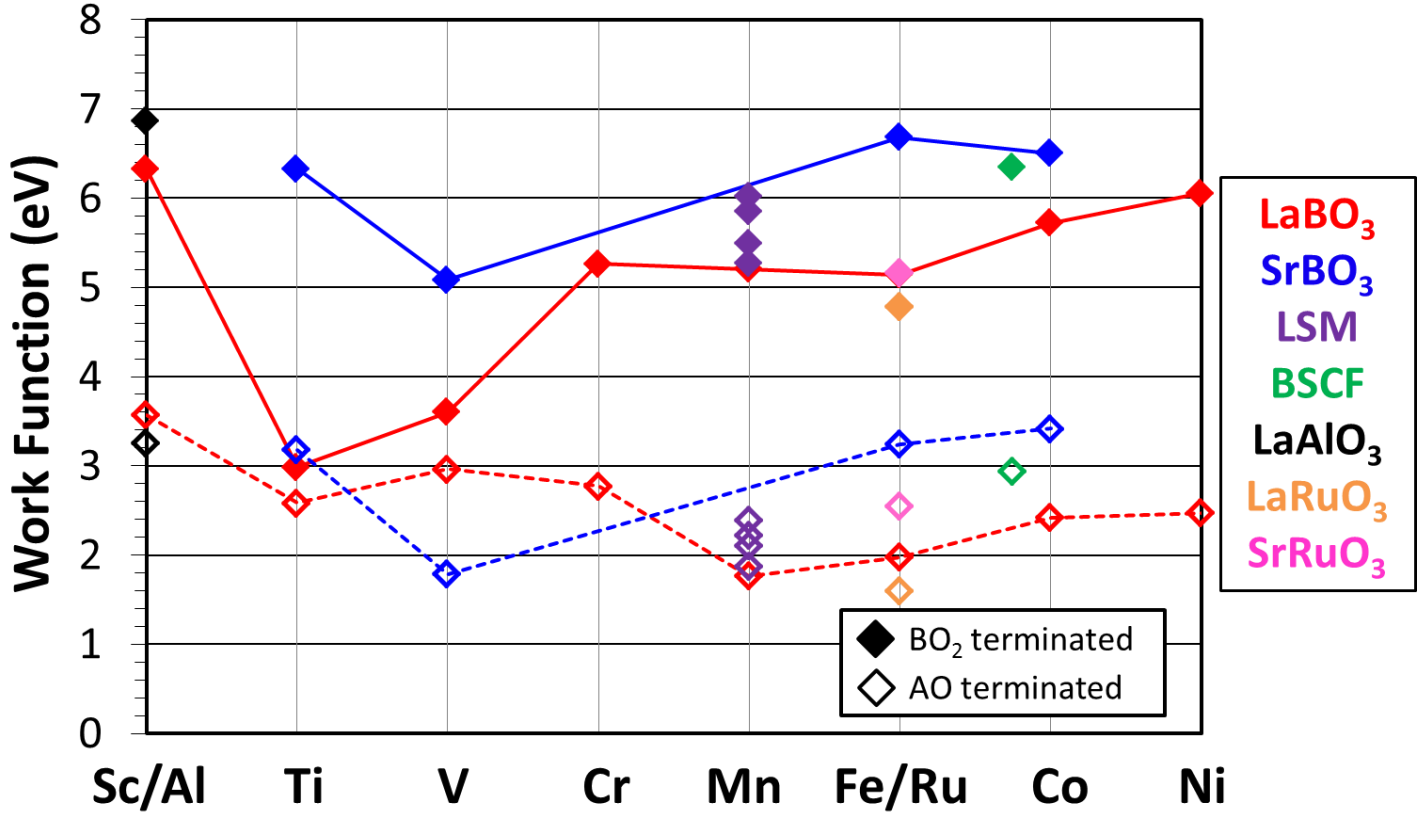

**Figure 2:** Trend of (001) AO- and BO$_2$-terminated surface work functions for the 20 perovskite materials studied in this work as a function of B-site element across the periodic table. The solid (open) symbols connected with a solid (dashed) line are the BO$_2$ (AO) work functions, respectively. Red, blue, purple, black, green, orange and pink symbols signify the LaBO$_3$ series, SrBO$_3$ series, La$_{1-x}$Sr$_x$MnO$_3$, x= 0.0625, 0.125, 0.25, 0.375 (LSM), LaAlO$_3$, Ba$_{0.5}$Sr$_{0.5}$Co$_{0.75}$Fe$_{0.25}$O$_3$ (BSCF), LaRuO$_3$ and SrRuO$_3$ materials, respectively. Note that the data points of BO$_2$ LaFeO$_3$ and SrRuO$_3$ work functions lie on top of each other.

**Table 1:** Summary of HSE calculated work functions for all (001) surfaces of ABO$_3$ materials considered in this study. Also listed are the ionicities (see text for definition) of the O-bonds for each material, which were calculated from Bader charge analysis of the bulk materials. A decrease in bonding ionicity is indicative of greater hybridization of the B 3*d* bands and O 2*p* bands. The range of reported experimental work functions for select materials is also provided. Additional discussion comparing calculated and experimental work functions is available in the **SI**.

| Material | AO WF (eV) | BO$_2$ WF (eV) | O bond ionicity | Experimental WF range (eV)** |
|---|---|---|---|---|
| LaScO$_3$ | 3.57 | 6.32 | 0.716 | -- |
| LaTiO$_3$ | 2.59 | 2.99 | 0.684 | -- |
| LaVO$_3$ | 2.97 | 3.60 | 0.664 | -- |
| LaCrO$_3$ | 2.77 | 5.27 | 0.665 | 4.3[32] |

| Material | | | | |
|---|---|---|---|---|
| LaMnO$_3$ | 1.76 | 5.21 | 0.653 | 4.5-5.1[30] [31] [32] |
| LaFeO$_3$ | 1.98 | 5.14 | 0.637 | 4.6[32] |
| LaCoO$_3$ | 2.42 | 5.73 | 0.599 | 5.0[32] |
| LaNiO$_3$ | 2.47 | 6.06 | 0.559 | 5.5[32] |
| LaAlO$_3$ | 3.25 | 6.87 | 0.879 | 2.2-4.2[2] |
| LaRuO$_3$ | 1.60 | 4.78 | 0.597 | -- |
| SrTiO$_3$ | 3.18 | 6.33 | 0.655 | 2.6-4.9[37] [38] |
| SrVO$_3$ | 1.86 | 5.09 | 0.601 | -- |
| SrFeO$_3$ | 3.24 | 6.68 | 0.556 | -- |
| SrCoO$_3$ | 3.42 | 6.51 | 0.537 | -- |
| SrRuO$_3$ | 2.55 | 5.16 | 0.614 | -- |
| Ba$_{0.5}$Sr$_{0.5}$Co$_{0.75}$Fe$_{0.25}$O$_3$ | 2.94 | 6.35 | 0.540 | -- |
| La$_{0.9375}$Sr$_{0.0625}$MnO$_3$ | 2.11 | 5.28 | 0.650 | -- |
| La$_{0.875}$Sr$_{0.125}$MnO$_3$ | 2.23 | 5.49 | 0.647 | -- |
| La$_{0.75}$Sr$_{0.25}$MnO$_3$ | 1.87 | 6.02 | 0.643 | 5.7[39] |
| La$_{0.625}$Sr$_{0.375}$MnO$_3$ | 2.39 | 5.85 | 0.637 | 4.7-4.9[39] [38] |

\*\* The experimental work function values reported here were measured using a variety of techniques on different sample types (e.g., thin film, polycrystalline compacts), which can render one-to-one comparisons between calculated and experimental work functions difficult. A detailed discussion comparing experimental and calculated work functions and possible reasons for discrepancy is presented in the **SI**.

Doping Sr into LaMnO$_3$ to produce LSM resulted in an increase of the AO and BO$_2$ work functions for all Sr concentrations relative to undoped LaMnO$_3$. When replacing La$^{3+}$ with Sr$^{2+}$, the system becomes more oxidized, i.e. it becomes hole-doped. This is evident from the work function data for the LSM series, where increasing the A-site Sr content tends to increase the work function of both surfaces and decrease the ionicity of the O-bonding. The fact that all BO$_2$ and most AO SrBO$_3$ material work functions are higher than their corresponding LaBO$_3$ work functions (with the exception of AO-terminated SrVO$_3$) is consistent with the intuition that doping Sr in place of La should raise the work function of the perovskite. We note that the increase in the work function of (La,Sr)MnO$_3$ is not monotonic with increasing Sr content. This lack of monotonic behavior is most likely a result of the specific Sr ordering chosen, and more work is needed to better understand work function variations in the disordered solid solution structure of La and Sr in LSM. Interestingly, BSCF has a lower work function than both SrFeO$_3$ and SrCoO$_3$, suggesting that doping Ba in place of Sr results in a lowering of the work function for Sr-based perovskites. The effect of Ba doping on the SrVO$_3$ work functions will be examined further in **Section 4**. The AO-terminations of SrVO$_3$, LaMnO$_3$ and LaRuO$_3$ have the lowest calculated work functions of 1.86 eV and 1.76 eV, and 1.60 eV respectively. By virtue of these low work functions, SrVO$_3$, LaMnO$_3$, and LaRuO$_3$ may be suitable candidates for low-work-function, electron-emission cathode materials. Of these candidate materials, SrVO$_3$ is a particularly promising material to explore for the emission applications discussed in the introduction section by virtue of its low work function, metallic conductivity, ability to be synthesized as both a bulk powder[40, 41] and (001)-oriented thin film,[42] and structural stability at high temperatures.[40, 41, 43] We therefore study SrVO$_3$ in more detail in **Section 4**.

### 2.2. 2*p*-band center as an electronic structure descriptor

Having demonstrated qualitative work function trends with changing A- and B-site composition for the ABO$_3$ materials investigated here, we turned our focus to developing a greater understanding of the physics governing the value of the work function in these perovskite materials. To accomplish this, we used the O 2*p*-band center (see **Section 2** of the **SI** for details) as an electronic structure descriptor, as this variable has proved useful for correlating with a number of perovskite properties.[6, 44-46] The B-site cation 3*d*-band center and the La/Sr A-site band centers (both calculated with respect to E$_{Fermi}$) were also investigated as possible descriptors. However, no useful physical trends emerged from their analysis. Therefore, we focused on the bulk O 2*p*-band center.

**Figure 3** demonstrates the relationship between the calculated (001) work functions and the value of the bulk O 2*p*-band center. **Figure 3A** (**Figure 3B**) is a plot of BO$_2$ work function (AO work function) as a function of the O 2*p*-band center energy. In both plots, the blue symbols refer to insulating perovskites while red symbols refer to metallic perovskites. In the present case, "insulating" refers to any material we calculated to have a finite bulk and surface band gap, whether due to band-insulating or Mott-Hubbard insulating behavior. The materials which compose the set of insulating perovskites are: LaScO$_3$, LaTiO$_3$, LaVO$_3$, LaCrO$_3$, SrTiO$_3$ and LaAlO$_3$. The remaining perovskite materials are referred to as "metallic" perovskites. Although the bulk ground states of some of these materials, for example LaMnO$_3$ and LaFeO$_3$, are also insulating, the ferromagnetic near-surface electronic structure is metallic.[47] Our inclusion of these materials in the category of "metallic" perovskites is appropriate as these materials demonstrate fundamentally different electronic structure behavior than prototypical Mott-Hubbard insulators such as LaTiO$_3$ and LaVO$_3$ near their surfaces. The influence of band positions and the O 2$p$-band center relationship of **Figure 3** are discussed in **Section 3.1**.

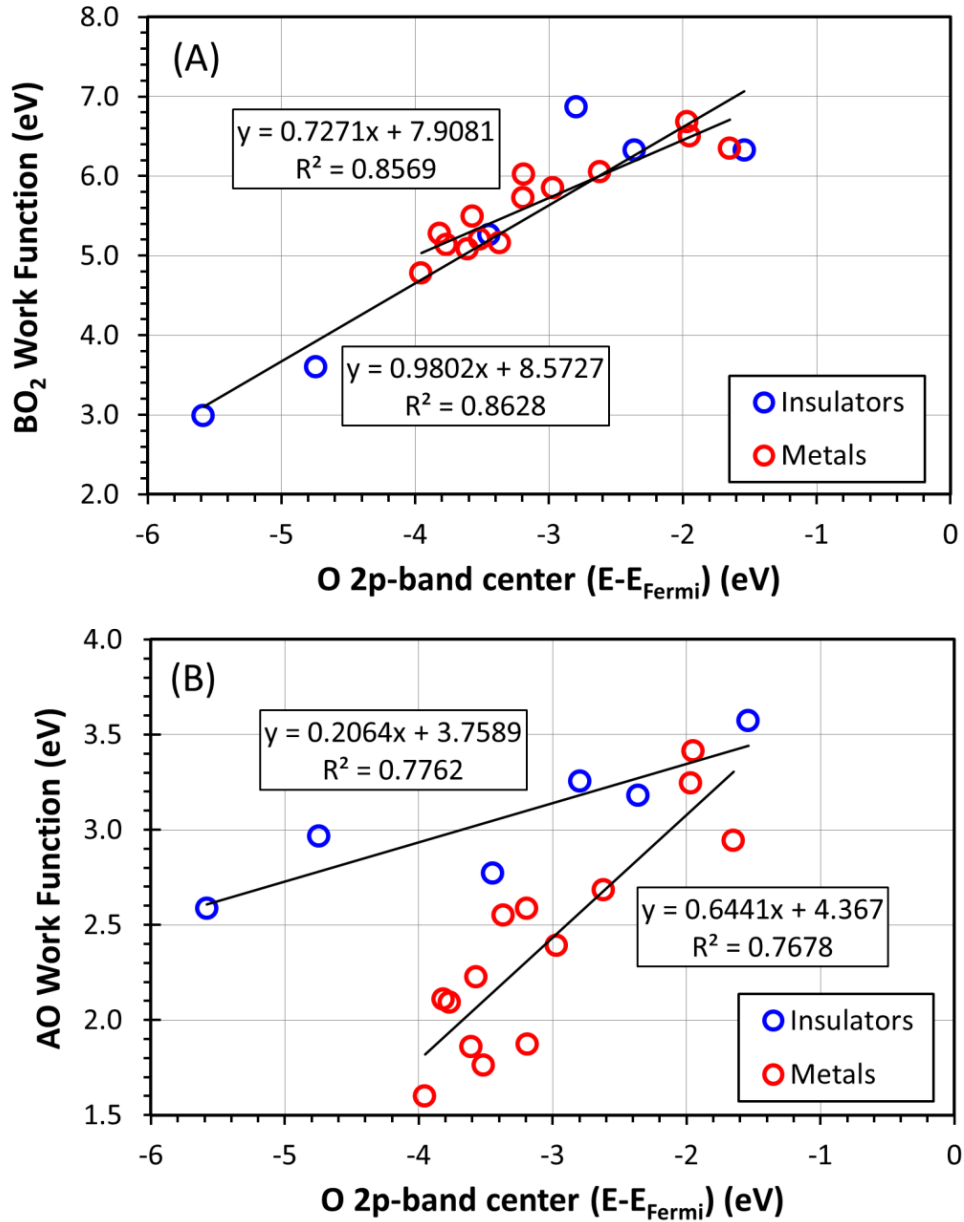

**Figure 3:** Plots of calculated work functions for the BO$_2$ (A) and AO-terminated surfaces (B) of ABO$_3$ materials as a function of the O 2$p$-band center of bulk ABO$_3$ materials. In both plots, the blue symbols represent insulating perovskites while the red symbols represent metallic perovskites. In (A) and (B) there is a semi-quantitative linear relationship between the work function and the O 2$p$-band center.

# 3. Discussion: ABO$_3$ work function data and trends

## 3.1. ABO$_3$ work function trends

Perhaps the most striking feature of the calculated work functions in **Figure 2** is that the AO surfaces generally have dramatically lower work function values than BO$_2$ surfaces in all cases. Qualitatively, this can be understood in terms of the surface dipoles. The alternating layers of the (001) orientation are AO/BO$_2$/AO/BO$_2$, which, when considering formal charges for $A^{3+}B^{3+}O_3^{6-}$ compounds, alternates +/-/+/-. A positive surface dipole is a dipole with an outwardly pointing positive charge, while a negative surface dipole has an outwardly pointing negative charge. Thus we see that the AO termination forms a positive surface dipole that decreases the work function, and the BO$_2$ surface forms a negative surface dipole that increases the work function. It is evident that the same trend of high BO$_2$ work functions and low AO work functions also occurs for the SrBO$_3$ (and BSCF) materials. Considering formal charges for these $A^{2+}B^{4+}O_3^{6-}$ compounds, one should expect no difference in surface dipole as the $A^{2+}O^{2-}$ and $B^{4+}O_2^{2-}$ surfaces both sum to zero charge. However, a number of previous studies have shown that surfaces of $A^{2+}B^{4+}O_3^{6-}$ compounds are in fact polar due to enrichment or deficiency of electrons relative to their formal valence. [47, 48]

The trend of increasing BO$_2$ work function when proceeding from left to right on the periodic table along the 3$d$ row as shown in **Figure 2** can be understood in terms of the transition metal electronegativities. When proceeding from Ti to Ni, the electronegativity of the transition metal ion is increasing as the 3$d$ band fills. As a result, when proceeding from Ti to Ni, the 3$d$ bands fill with more electrons, the 3$d$ bands shift lower in energy, and the work function increases. For the materials LaScO$_3$, LaAlO$_3$ and SrTiO$_3$, the 3$d$ bands are nearly empty and these materials behave as band insulators. Interestingly, these band insulator materials have nearly the same BO$_2$ and AO work function values within a few tenths of an eV. For the case of these band insulators, the absence of 3$d$ electrons means the Fermi level resides at the top of the O 2$p$ band. Thus, for materials with no 3$d$ electrons, it is the O 2$p$ band which sets the value of the work function.

From **Table 1**, it is evident that as the 3*d* bands fill, the bonding ionicity decreases, which implies that the B 3*d* and O 2*p* bands are becoming more hybridized. In addition, for the SrBO$_3$ materials where the B element is in the 4+ oxidation state, the bond ionicities are lower and thus the B 3*d* and O 2*p* bands are more hybridized than the corresponding LaBO$_3$ systems where the B element is in the 3+ oxidation state. These trends of B 3*d* - O 2*p* band hybridization are consistent with a joint experimental and computational work by Suntivich and coworkers that showed how B 3*d* - O 2*p* band hybridization changes as a function of 3*d* band filling using O K-edge X-ray absorption and DFT calculations on a series of perovskite and Ruddlesden-Popper materials.[49] The increased B 3*d* - O 2*p* band hybridization means there is greater overlap of the B 3*d* and O 2*p* bands, and the O 2*p* band center becomes closer to E$_{Fermi}$. The above trends in hybridization with 3*d* band filling illustrate that materials with higher 3*d* band filling will have increased band hybridization. This increased band hybridization will result in O 2*p* bands that are closer to E$_{Fermi}$ , which will result in higher BO$_2$ work functions. Based on this discussion, if one were interested in creating a perovskite with a higher work function, one strategy would be to increase the band hybridization by doping the A-site with alkaline earths to further oxidize the B-site, for example Sr doping in LaMnO$_3$ to create LSM or Sr and Ba doping in La(Co, Fe)O$_3$ to create BSCF.

Both plots in **Figure 3** show a linear trend of the calculated work function versus the bulk O 2*p*-band center, although the trend is more consistently linear (e.g., it has a higher $R^2$ value) in the case of BO$_2$ work functions. In general, these results demonstrate that the bulk O 2*p*-band center provides an approximate predictor of the work function. Interestingly, in **Figure 3A**, the slopes of the BO$_2$ work function versus O 2*p*-band center are approximately 0.7 and 1, which is generally close to one, while in **Figure 3B** the slopes of the AO work functions are approximately 0.2 and 0.6. Since the work function is controlled by a combination of the energy band positions and the magnitude of the surface dipole, the slope of BO$_2$ work function versus O 2*p*-band center being nearly equal to one implies that the BO$_2$ work function changes are dominated by the band positions and are thus relatively insensitive to the magnitude of the surface dipole. In the case of the AO work functions, where the work function doesn't change as rapidly in proportion to the movement of the O 2*p*-band center, we conclude that the AO work functions are dominated by surface dipoles and are relatively insensitive to the band positions.

Unfortunately, direct comparison with experimental values is difficult as there are no experiments that measure the work function of a specific perovskite surface termination, as we are calculating here. However, because DFT has been shown to accurately reproduce work functions of metal surfaces[21] and the ability of HSE to accurately reproduce the electronic structure of these materials,[19, 20] it is reasonable to expect that the calculated work functions are within a few tenths of an eV of the true work function for the surface being modeled. Some additional errors are introduced due to the use of idealized surfaces without defects or reconstructions, but we expect those effects to also be within a couple of tenths of an eV for most surfaces, as discussed in **Section 3** of the **SI**. Encouragingly, our calculated work function values follow the same compositional trend as the surface averaged work function values for La(Cr, Mn, Fe, Co, Ni)O$_3$ materials obtained with X-ray absorption and emission spectroscopy.[32] Additional discussion regarding causes of quantitative differences between experimental and calculated work functions is provided in **Section 3** of the **SI**.

### 3.2. Density of States Pictures of Work Function Trends

As discussed in **Section 3.1**, the band positions are the dominant contribution setting the value of the BO$_2$ work functions. Here, we illustrate the connection between the band structure and work function using schematic density of states diagrams. **Figure 4** is a density of states schematic that illustrates the trend of BO$_2$ work functions from **Figure 2** by comparing the density of states of an insulating material with an empty 3$d$ band and high ionicity (small amount of B 3$d$- O 2$p$ hybridization) such as LaScO$_3$ (**Figure 4A**), a less ionic material (large amount of B 3$d$- O 2$p$ hybridization) with half or mostly filled 3$d$ band such as LaNiO$_3$ (**Figure 4B**) and a metallic, medium ionicity material (medium amount of B 3$d$- O 2$p$ hybridization) with a minimally occupied 3$d$ band such as SrVO$_3$ (**Figure 4C**). The vacuum level, Fermi level and O 2$p$-band center are denoted as E$_{vac}$, E$_{Fermi}$, and $\bar{O}_{2p}(E)$, respectively. Following the convention of DFT calculations, the position of E$_{Fermi}$ is at the energy of the highest filled electronic state. The O 2$p$ states are shown in red and the B 3$d$ states are shown in blue. The states that are shaded are filled states. In **Figure 4**, we made the simplifying approximation that for a fixed surface dipole the O 2$p$ bands remain at a fixed energy level relative to vacuum. Maintaining a constant level of the O 2$p$ band provides us with a straightforward and intuitive way to demonstrate how the work function varies with the band positions and associated properties such as bond ionicity/hybridization and also how the value of the O 2$p$-band center ($x$-axis in **Figure 3**) physically relates to our calculated work function values. While our calculated results are most consistent with a fixed position for the O 2$p$-band center relative to vacuum under a fixed surface dipole, it is difficult to prove that this rigorously occurs, and some movement of O 2$p$-band center relative to vacuum is certainly possible between materials. We note that results of X-ray absorption and emission measurements of La(Cr, Mn, Fe, Co, Ni)O$_3$ polycrystalline samples show that the occupied O 2$p$ states may move relative to the vacuum level by approximately +/- 1 eV for varying B-site composition.[32] When we averaged our calculated bulk O 2$p$-band centers relative to the vacuum level for the BO$_2$- and AO-terminated surfaces we found the standard deviation of O 2$p$-band center was 0.4 and 1 eV, respectively, which is qualitatively consistent with the spectroscopy results of Ref. [32]. However, these measured and calculated changes mix both movement of the O 2$p$-band center and changes in the surface dipoles, and we believe that for fixed dipoles the O 2$p$-band center position relative to vacuum may be quite stable. The Δ values indicate the energy difference between the O 2$p$-band center and E$_{Fermi}$, equivalent to the $x$-axis of **Figure 3**. In **Figure 4A**, the insulating perovskite with empty 3$d$ band has very deep O 2$p$ bands which results in a deep E$_{Fermi}$ (relative to the vacuum level), an O 2$p$-band center close to E$_{Fermi}$ and high work function. Because we use the DFT convention with E$_{Fermi}$ located at the valence band maximum, the diagram in **Figure 4A** shows the p-type limit of the work function (i.e., ionization potential) for an insulating perovskite. In **Figure 4B**, the perovskite with partially filled 3$d$ band has a large amount of band hybridization (i.e. lower ionicity) which results in higher occupied electron energy states, an O 2$p$-band center further from E$_{Fermi}$ compared to **Figure 4A**, and a slightly lower work function. In **Figure 4C**, the metallic perovskite with minimally filled 3$d$ band has less band hybridization than the case in **Figure 4B**, which results in an occupied portion of the B 3*d* band that is more empty, less hybridized and is higher in energy. Since the occupied portion of the B 3*d* band is higher in energy, $E_{Fermi}$ is also higher. Overall, this leads to an O 2*p*-band center that is further from $E_{Fermi}$ and a lower work function.

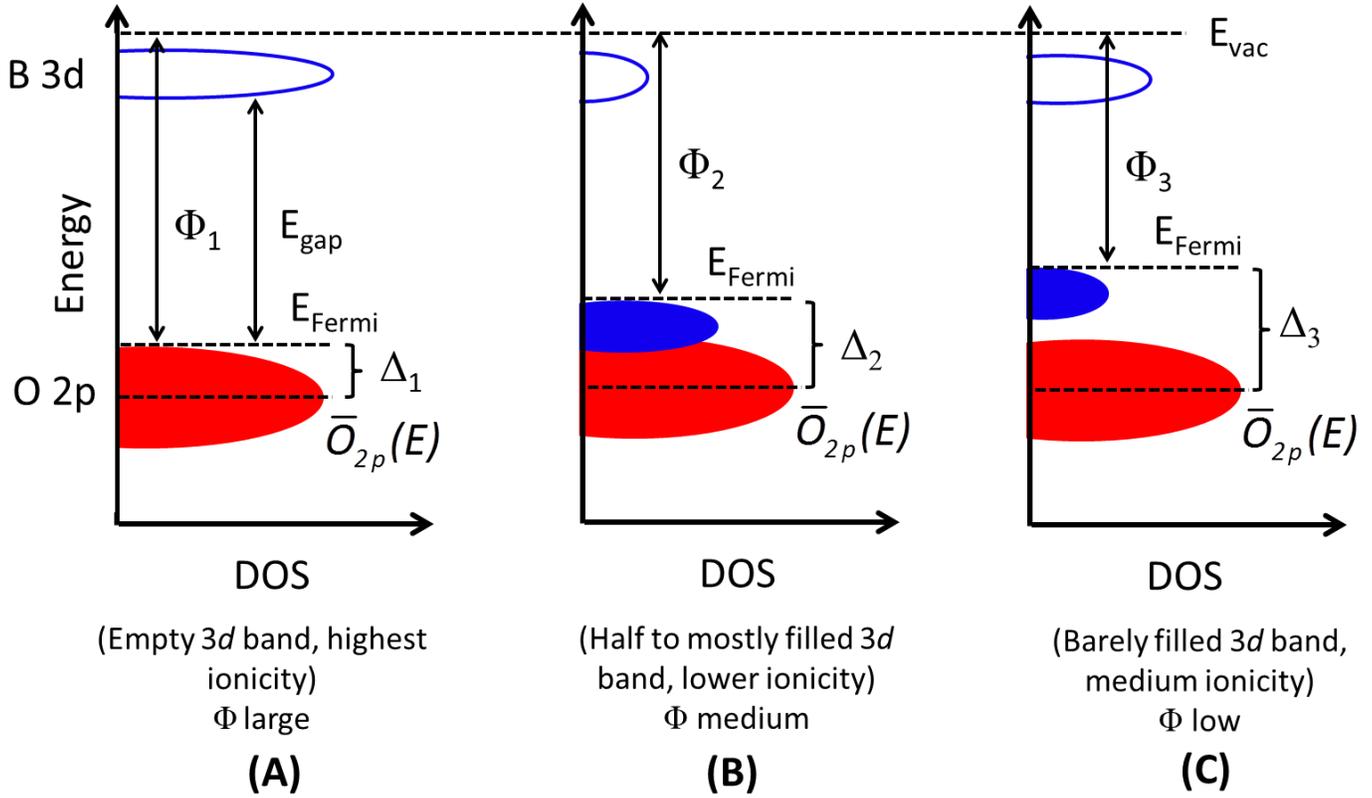

**Figure 4:** Schematic density of states plots for (A) insulating perovskite with empty 3*d* band such as LaScO$_3$, (B) perovskite with partially or mostly filled 3*d* band such as LaNiO$_3$, and (C) metallic perovskite with minimally filled 3*d* band such as SrVO$_3$. The red regions denote O 2*p* states while the blue regions denote B 3*d* states. Shaded regions indicate filled states while unshaded regions denote empty states. The labels and symbols are defined in the main text. The case in plot (C) of a material with minimally filled 3*d* band results in an O 2*p*-band center furthest below $E_{Fermi}$ and a low work function. The $\Delta$ values are defined as the difference between the O 2*p*-band center and $E_{Fermi}$, equivalent to the *x*-axis of **Figure 3**.

The insensitivity of the BO$_2$ work functions to surface dipoles can be understood qualitatively by considering the origin and path of emitted electrons. These densities of states in **Figure 4** show that for materials containing 3*d* electrons the states at $E_{Fermi}$ are dominated by hybridized B 3*d* and O 2*p* states. Therefore, we can think of the emitted electrons as emerging from the BO$_2$ layers. Thus, the electrons being emitted from the $BO_2$ surface are already at the surface and can be directly emitted into vacuum. Recall that, in contrast to the $BO_2$ work function, the AO work functions are dominated by large surface dipoles. Emission from the AO surface involves electrons moving from the subsurface $BO_2$ layer through the AO layer and being emitted. This difference in pathway makes the $BO_2$ surface work function largely insensitive to the surface dipole but the AO surface work function very sensitive to the surface dipole. This explanation of a surface-emitting electron experiencing a large, work function-lowering surface dipole via the AO-surface is shown in **Figure 5**.

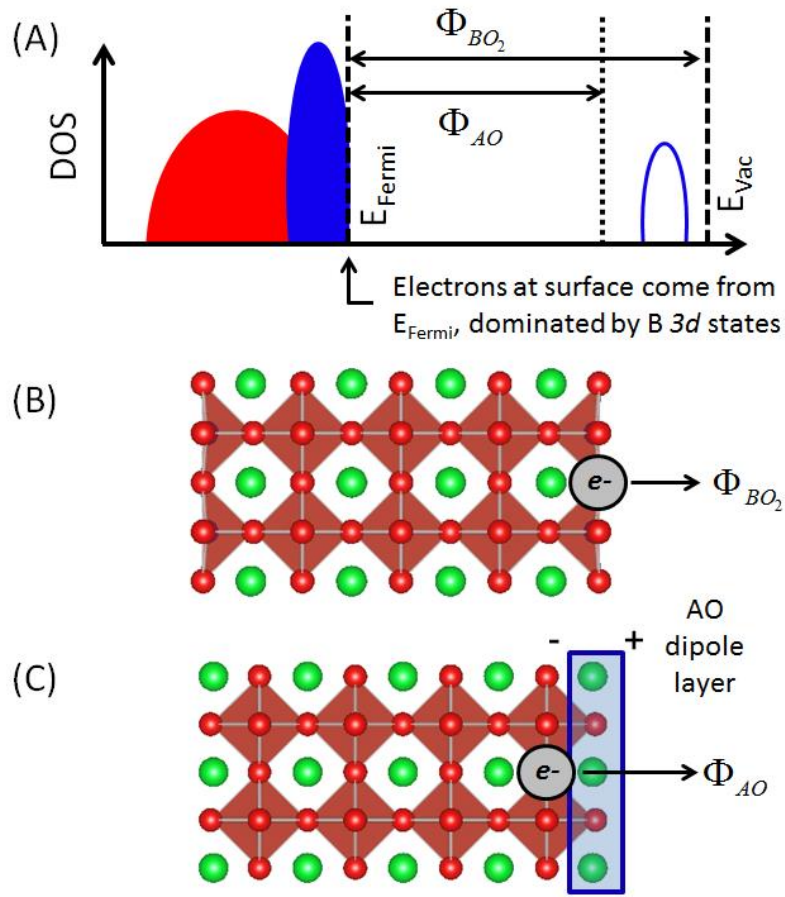

**Figure 5:** Schematic demonstrating why the $BO_2$ work functions are insensitive to surface dipoles, and the AO work functions are dominated by surface dipoles. (A) Density of states representative of a transition metal perovskite with partially occupied $3d$ band. The colors and labels are the same as **Figure 4**. The Fermi level is dominated by B $3d$ states for both the AO- and $BO_2$-terminated surfaces. Electrons emitting from the surfaces of these materials will originate at the Fermi level, which are predominantly from the $BO_2$ layers. In (B), the emission of surface electrons from the $BO_2$ surface originates from the terminating surface layer. The $BO_2$ work function is dominated by the band positions relative to vacuum, and relatively insensitive to the surface dipoles. In (C), the emission of a surface electron again originates from the $BO_2$ layer, which is now present in the subsurface layer. The terminating surface layer is now an AO-plane, which does not appreciably contribute to the density of states near the Fermi level. Here, the AO surface acts as a large dipole layer, lowering the AO work function. In this way, the work function of the AO surface is dominated by surface dipoles, and is relatively insensitive to the band positions.

From the above discussions we can summarize our understanding of the trend in O 2$p$ band with the work function as follows: The location of the O 2$p$ band relative to $E_{Fermi}$ is highly dependent on the number of 3$d$ electrons in the system and the hybridization between the B 3$d$ levels and O 2$p$ levels. When proceeding from Ti through Ni, more 3$d$ electrons are added to the system, the bond hybridization increases, the 3$d$ bands fill and move lower in energy, and thus $E_{Fermi}$ is lower in energy and closer to the O 2$p$-band center. Because $E_{Fermi}$ is lower in energy, the work function of BO$_2$ surfaces increases as more 3$d$ electrons are added. Furthermore, for the same B-site transition metal element, if the B-site is more oxidized (e.g. comparing Co$^{3+}$ in LaCoO$_3$ with Co$^{4+}$ in SrCoO$_3$), the material containing the more oxidized transition metal will exhibit greater hybridization between the B 3$d$ and O 2$p$ bands, thus resulting in higher work functions. From **Table 1** and **Figure 2**, one can see that all SrBO$_3$ materials have higher work functions than their analogous LaBO$_3$ materials, except for AO-terminated SrVO$_3$. These hybridization trends with 3$d$ electron filling are consistent with experimental and computational findings of Suntivich and coworkers.[49] Broadly, the band structure progression shown in **Figure 4** is a close representation of how the BO$_2$ work function changes with composition and 3$d$ band filling. In the case of the AO work function, the portion of the work function due to surface dipoles is strong enough such that it overwhelms the band physics contributions of **Figure 4**. This yields a weaker relationship between work function and O 2$p$-band center (slope significantly less than one, see **Figure 3B**).

## 4. Results and Discussion: SrVO$_3$ as a low work function, metallic perovskite

Our earlier analysis in **Section 2.1** has demonstrated that of the 18 perovskite materials considered here, SrVO$_3$ is one of the most promising materials for electron emitting applications, in particular for high power electron beam devices used in defense, scientific research and communications, and as an electron-emitting layer in the renewable energy technology of photon-enhanced thermionic energy conversion devices. The metallic perovskite SrVO$_3$ has been successfully synthesized both as a bulk polycrystalline powder[40, 41] and as a controlled (001)-oriented thin film grown with MBE.[42] SrVO$_3$ possesses a very high conductivity of about 10$^5$ Ω$^{-1}$cm$^{-1}$ at room temperature, higher than SrRuO$_3$ (a prototypical metallic perovskite) and on par with elemental metals such as Pt.[42] SrVO$_3$ maintains its structural stability even up to high temperatures of 1300 ºC and under reducing conditions during annealing with an H$_2$/N$_2$ or H$_2$/Ar gas atmosphere.[40, 41, 43] Because perovskites are receptive to compositional modification, there are opportunities with doping SrVO$_3$ to lower its work function further. In this section, we consider alkaline earth metal doping in SrVO$_3$. We also consider the pristine (011) and (111) surface terminations to ascertain the full work function range of SrVO$_3$ and also obtain a more quantitative understanding of which surface terminations should be stable (and thus present in the highest quantity) in a real device. In addition, we consider the effect of surface segregation in SrVO$_3$ as a number of studies have suggested that A-site alloyed perovskites can show significant cation segregation.[47, 50-58]

**Figure 6** contains the surface structures of (011) and (111) terminated SrVO$_3$. From **Figure 6A**, the (011) termination can either be O-terminated or ABO-terminated. **Figure 6B** and **Figure 6C** show symmetric (111) surfaces that are B-terminated (**Figure 6B**) and AO$_3$-terminated (**Figure 6C**). The work functions and surface energies for these surface terminations (as well as surface energies for (001) surfaces) were calculated and are tabulated below in **Table 2**.

**Table 2:** Tabulated values of calculated work functions and surface energies for different SrVO$_3$ surface terminations. The work functions of (001) surfaces are repeated from **Table 1** for clarity.

| Termination | Work Function (eV) | Surface Energy (eV/Å$^2$) |
|---|---|---|
| (001) | 1.86 (AO), 5.09 (BO$_2$) | 0.052 (AO/BO$_2$ average) |
| (011) | 2.32 (ABO), 7.23 (O) | 0.094 (O/ABO average) |
| (111) | 2.78 (B), 4.68 (AO$_3$) | 0.078 (B/AO$_3$ average) |

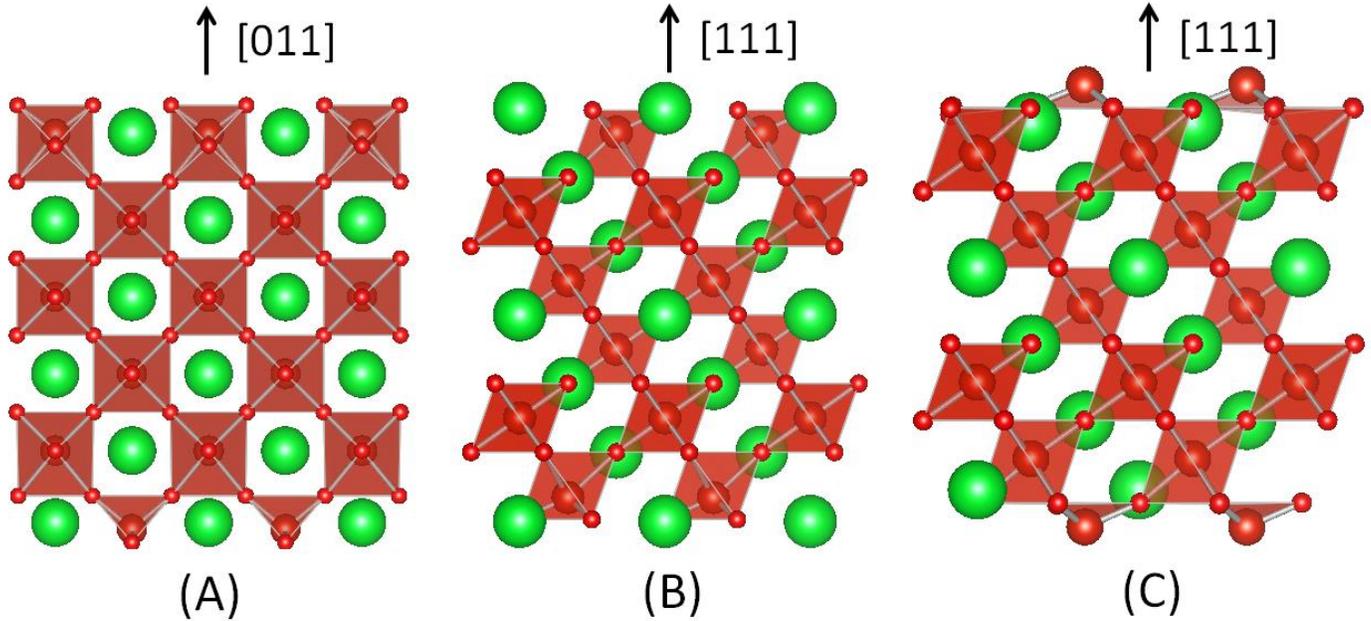

**Figure 6:** SrVO$_3$ surface slabs of (011) and (111) orientations. (A) The (011) orientation, whereby the top surface is O-terminated and the bottom surface is ABO-terminated. (B) The (111) orientation, with both surfaces terminated as AO$_3$. (C) The (111) orientation, now with both surfaces B-terminated. The large green spheres are Sr, medium-sized red spheres are V (in the middle of the octahedra), and the small red spheres are O.

From **Table 2**, one can see that the pristine (001) surfaces have a lower surface energy and thus more stable than (011) and (111) surfaces, consistent with previous DFT studies.[47, 51, 59] Recent experimental[60-64] and computational studies[57, 64, 65] show that numerous perovskite materials exhibit segregation of alkaline earth elements such as Sr and Ba. From our current calculations, the overall order of stability is: $\gamma(001) < \gamma(111) < \gamma(011)$. The ABO-terminated (011) surface has a reasonably low work function of 2.32 eV, but overall the (011) and (111) surfaces possess higher work functions than AO-terminated (001). The fact that the (001) terminations of SrVO$_3$ are predicted to be the stable terminations, together with the fact that AO-terminated (001) SrVO$_3$ exhibits the lowest work function of the surfaces explored here, further reinforces the choice of SrVO$_3$ as a potentially new low work function material.

We now turn to examining the effect of doping the alkaline earth metals Mg, Ca and Ba in SrVO$_3$. From **Figure 2**, it was suggested from comparing the work function values of SrFeO$_3$, SrCoO$_3$, and BSCF that doping Ba onto the A-site of Sr-based perovskites may result in a lowering of the work function. Here, we focus solely on the AO-terminated (001) surface of SrVO$_3$ since this is the low work function surface termination of interest. The AO-terminated (001) surface was simulated with concentrations of 25%, 50% and 100% (this is equivalent to replacing one SrO row with a (Mg, Ca, Ba)O row) site fraction Mg, Ca and Ba on the surface of the AO (001) slab (see **Figure 7C**). As shown in **Figure 7**, it was found that surface doping of Mg and Ca raised the work function for all concentrations, while doping Ba lowered the work function for all concentrations. In particular, a site fraction of 100% Ba on the surface resulted in a very low work function of just 1.07 eV.

To better understand the role of Ba doping in lowering the work function (i.e. bulk doping versus surface dipole formation), we also simulated a full layer of BaO in place of SrO in the middle of the AO (001) slab. It was found that placement of Ba in the middle of the slab resulted in a barely increased work function of 1.90 eV, which is 0.04 eV higher than pure $SrVO_3$. However, placement of the Ba in the top surface layer resulted in a significant lowering of the work function down to 1.07 eV, which is 0.79 eV lower than pure $SrVO_3$. This indicates that the work function lowering from Ba doping is due entirely to altering the surface dipole, rather than altering the Fermi level. By comparing the atomic positions of a pristine $SrVO_3$ surface and $SrVO_3$ with Ba in the surface layer, it is clear that the bond lengths between Ba and sub-surface O (the O in the $BO_2$ layer beneath the surface) is about 0.2 Å longer than the bond length between Sr and the same sub-surface O. This longer bond length is most likely the result of the larger ionic radius of Ba (1.75 Å) over Sr (1.58 Å). [66] This bond lengthening is expected to increase the size of the dipole for Ba at the surface in a direction that will lower the work function compared to Sr, and this bond lengthening is likely a major reason for the work function change with Ba doping. The work function reduction of 0.79 eV amounts to a surface dipole change of approximately 0.26 eV-Å with the addition of a full Ba surface layer, which can be obtained directly from VASP simulations and is also calculable using the Helmholtz equation.[21, 22]

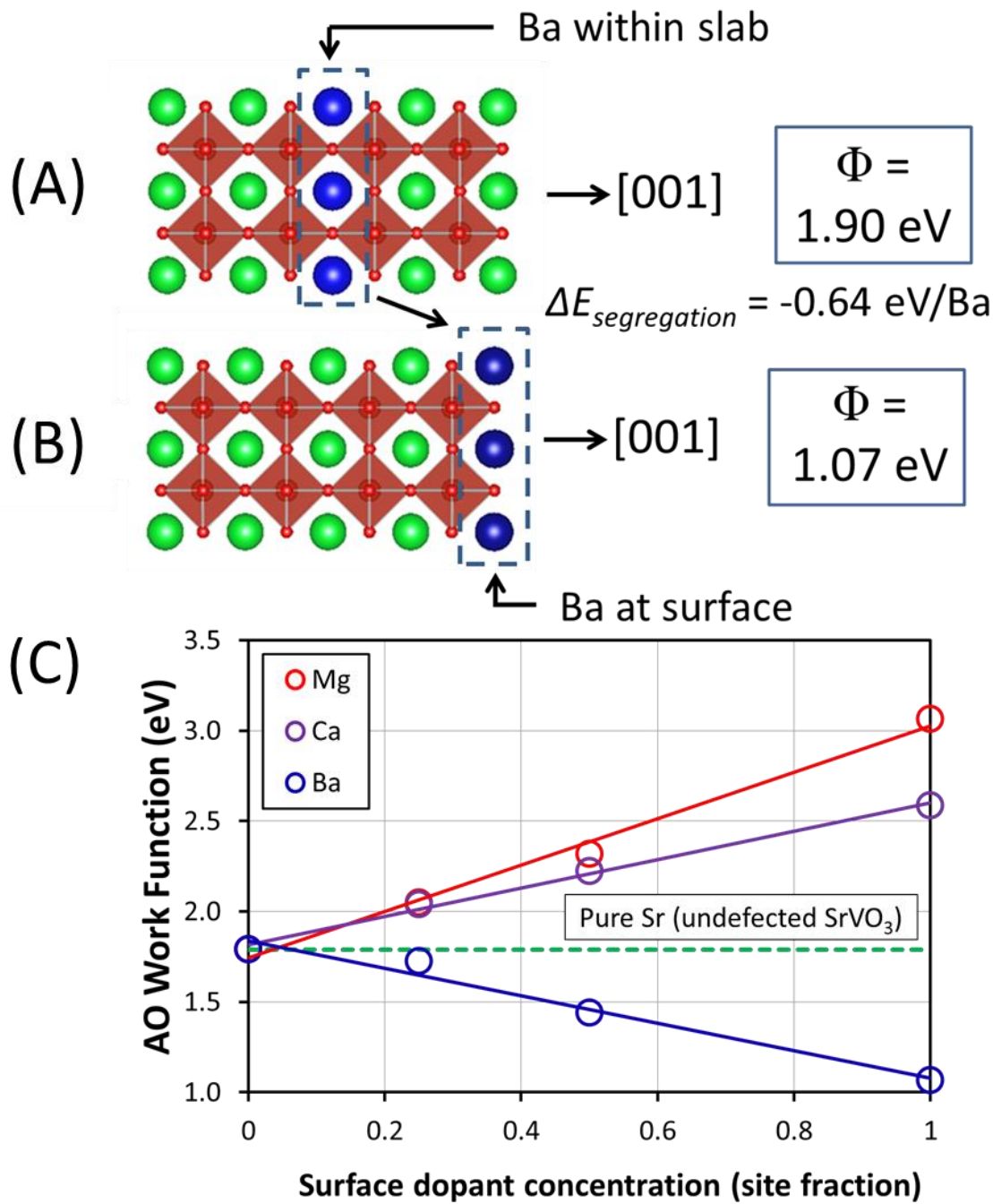

**Figure 7:** Simulated AO-terminated (001) SrVO$_3$ surfaces with a single SrO layer replaced by AO (A=Mg, Ca, Ba), with a focus on Ba. (A) Ba doping in the middle of the surface slab, resulting in a work function of 1.90 eV. (B) Ba doping at the surface of the slab, resulting in an extremely low work function of 1.07 eV. The Ba segregation energy is calculated to be -0.64 eV/Ba, and indicates that Ba will preferentially segregate to the surface. The large green spheres are Sr, the large blue spheres are Ba, medium-sized red spheres are V (in the middle of the octahedra), and the small red spheres are O. The plot in (C) shows how the calculated AO-terminated SrVO$_3$ work function changes when the top surface layer is alloyed with Mg, Ca, and Ba for different concentrations. The only dopant expected to lower the work function is Ba.

Because $Ba^{2+}$ is a larger cation than $Sr^{2+}$, it was worth investigating whether cation segregation may occur in doped $SrVO_3$. As discussed previously, cation segregation has been observed in many perovskite materials.[50-57, 60-66] To ascertain if Ba segregation may occur in $SrVO_3$, we calculated the formation energy of substituting Ba in place of Sr for the two cases illustrated in **Figure 7**, and also calculated the segregation energy of dilute Ba (25% Ba substitution in the middle of the surface slab) to the surface of $SrVO_3$. The energy to substitute Sr for Ba, $\Delta E_{sub}$, was calculated using the equation $\Delta E_{sub} = E_{defected} - E_{perfect} - x(E_{BaO} - E_{SrO})$, where $E_{defected}$ is the total energy of the $SrVO_3$ surface slab with Ba substituting for Sr, $E_{perfect}$ is the energy of the undefected $SrVO_3$ slab, $x$ is the number of Ba substitutions (in this case $x=1$ Ba atom in our dilute calculation), and $E_{BaO}$ and $E_{SrO}$ are the total energies of rocksalt BaO and rocksalt SrO, respectively, which are taken as the reference states for Ba and Sr atoms. We found that the energy to substitute Ba for Sr in the middle of the $SrVO_3$ slab (**Figure 7A**) was 0.26 eV/Ba, while to substitute Ba for Sr on the surface (**Figure 7B**) was -0.38 eV/Ba. The energetic driving force for Ba surface segregation is just the difference of these energies, and is equal to -0.64 eV/Ba. Note that while the value of $\Delta E_{sub}$ is, in principle, dependent on temperature, pressure, and choice of reference state, the energy *difference* reported by calculating the segregation energy is the more physically insightful quantity, and its value is independent of the chosen reference state. The magnitude of this segregation energy is consistent with DFT calculations of cation surface segregation in other systems.[67] Therefore, if Ba is doped into $SrVO_3$, one may expect that over time Ba will diffuse to the surface and can dramatically lower the value of the work function. Analogous calculations for Mg and Ca doping indicate there is essentially no driving force (-0.07 eV/atom for Mg, -0.05 eV/atom for Ca) to segregate these species to the $SrVO_3$ surface compared to the Ba case. A combined experimental and DFT study of Ca, Sr, and Ba doping in (La, Sm)$MnO_3$ has suggested that cation segregation is a combination of both elastic (via lattice strain of mismatched cation sizes) and electrostatic effects attributed to the differing valences of alkaline earth and lanthanide elements as well as interaction with charged defects in doped $LaMnO_3$.[57] In our case of alkaline earth doping in $SrVO_3$ the predicted Ba cation segregation is presumably due primarily to lattice strain, as Mg, Ca, Sr and Ba are all 2+ cations and no charged defects or vacancies have been considered.

An important consideration of Ba doping in SrVO$_3$ is whether or not the surface-segregated Ba atoms are stable on the surface. To investigate this stability, we compared the adsorption energy of the Ba-O species present on the surface relative to bulk rocksalt BaO using standard GGA-based DFT methods for three cases: 1/4 monolayer Ba-O coverage on W(001) following Ref. [21], 7/8 monolayer Ba-O coverage on Sc$_2$O$_3$(011) following Ref. [22], and the present case of 1 monolayer Ba-O coverage on SrVO$_3$ (001). We use GGA-DFT methods here so that direct comparison with previous work can be made. These materials were chosen for comparison with SrVO$_3$ because W(001) with BaO is the dominant emitting surface of typical commercial thermionic cathode devices and Sc$_2$O$_3$(011) with BaO was found to be the most likely candidate for low work function surfaces in scandate cathode devices.[21, 22] We found that the adsorption energy (per Ba-O formula unit) for W(001), Sc$_2$O$_3$(011) and SrVO$_3$(001) are: 0.71 eV/Ba, -0.27 eV/Ba and -1.19 eV/Ba, respectively. Because the time to desorb an atom from a material surface scales exponentially with the adsorption energy, it is evident from the above calculations that at $T$ = 1000 K, which is an approximate temperature used in thermionic emission devices, Ba will reside on the SrVO$_3$(001) surface approximately 5 orders of magnitude longer than on Sc$_2$O$_3$(011), and approximately 9 orders of magnitude longer than on W(001). Overall, the surface-segregated Ba atoms in SrVO$_3$ are much more strongly bonded to the SrVO$_3$ surface than the volatile Ba-O surface dipole layers present in W- and Sc$_2$O$_3$-based electron sources. This may provide a method of obtaining an electron emission source that simultaneously exhibits an ultra-low work function of 1.07 eV and operating lifetime orders of magnitude longer than current dispenser cathode technologies.

As we have only considered a small representative set of perovskite materials in this study, it is possible that other low work function perovskite materials besides SrVO$_3$ exist. The O 2$p$-band center provides an approximate way to predict the work function of either the AO- or BO$_2$-terminated surface from strictly a bulk materials property. In general, surface supercell calculations are quite computationally expensive (especially with HSE functionals), while bulk calculations are many times faster as a result of fewer atoms per supercell and higher supercell symmetry. Thus, the correlation between bulk O 2$p$-band center and surface work function may enable fast, bulk materials screening of the O 2$p$-band center to predict work function values of perovskite alloys. Calculation of the bulk O 2$p$-band center is roughly a factor of 25 times faster than calculating the work function (a factor of 50 considering both the AO- and BO$_2$-terminated surfaces), and thus provides a useful estimate of a perovskite work function with comparatively minimal computational time. By high-throughput calculation of perovskite band gaps and O 2$p$-band centers one could screen for low work function materials.[68-70] In particular, any material that meets the conditions of zero (or near-zero) band gap and low O 2$p$-band center may warrant further investigation by way of surface calculations for quantitative work function values. Some preliminary high-throughput DFT screening using GGA+$U$ has indicated that perovskites within the family of (La, Pr, Y)(Ti, V)O$_3$ and SrVO$_3$ have deep O 2$p$ band centers and a partially filled 3$d$ band, and thus should have low work functions, consistent with the trends in **Figure 2**. While many of the (La, Pr, Y)(Ti, V)O$_3$ materials are Mott-Hubbard insulators and thus may not be sufficiently conducting for low work function cathode applications, further A-site alloying of alkaline earths and B-site alloying with other transition metals within this low O 2$p$-band composition space may potentially yield smaller (or zero) bandgap materials that are worth further investigation.

## 5. Conclusions

In this study, we have computationally explored 20 technologically relevant perovskite materials at idealized undefected compositions to understand how their chemistry influences the value of the work function, and then use this new understanding to search for a new, low work function material for electron emission applications. We have explained compositional trends in the work function using concepts of bonding ionicity, hybridization, band filling, and surface dipoles. We found that the bulk O 2$p$-band center can function as a semi-quantitative descriptor of the work function for both AO- and BO$_2$-terminated (001) perovskite surfaces. Our usage of the O 2$p$-band center descriptor has provided both further understanding of the work function physics for these materials and may enable fast computational screening of perovskite materials with a particular work function value. Broadly, the work function depends both on the bulk electronic band filling (E$_{Fermi}$ position) and the influence of surface dipoles. Based on our analysis, the value of the BO$_2$ surface work function is dominated by bulk band positions while the work function of the AO surfaces is dominated by surface dipoles. Understanding the work function trends in perovskites as well as the physics governing these trends (e.g., band filling versus surface dipoles) is pivotal for the rational design of perovskite-containing devices involving electron transport at interfaces or surfaces.

We have computationally predicted SrVO$_3$ to be a new, promising low work function material for electron emission applications such as high power microwave tubes, satellite communications, and possibly as an emissive layer for photon-enhanced thermionic energy converters. SrVO$_3$ is not only predicted to have a low work function, but has been experimentally shown to be stable in reducing environments and exhibit a high electrical conductivity. When doped with Ba, we have shown that Ba will preferentially segregate to the surface and result in an ultra-low work function down to nearly 1 eV. The Ba contained in the SrVO$_3$ surface-emitting layer is significantly more stable than the Ba-containing dipole layers used in conventional thermionic electron emission cathodes such as the B-type W dispenser and scandate cathodes, which should thus produce a stable, highly-emissive, long lifetime electron emitter.

## 6. Computational Methods

Perovskite structures generally form in the $Pm\bar{3}m$ (cubic, space group 221), *P4mm* (tetragonal, space group 99), *Pbnm/Pnma* (orthorhombic, space group 62), and $R\bar{3}c$ (rhombohedral, space group 167) symmetries.[71] In this study, we used idealized, undefected 2x2x2 pseudocubic structure lattice constants adopted from full relaxation of the ideal cubic ABO$_3$ symmetry ($Pm\bar{3}m$, SrTiO$_3$, SrVO$_3$, SrFeO$_3$, SrCoO$_3$, SrRuO$_3$, Ba$_{0.5}$Sr$_{0.5}$Co$_{0.75}$Fe$_{0.25}$O$_3$), orthorhombic symmetry (*Pbnm*, LaScO$_3$, LaTiO$_3$, LaVO$_3$, LaCrO$_3$, LaMnO$_3$, La$_{1-x}$Sr$_x$MnO$_3$, LaFeO$_3$, LaRuO$_3$), and rhombohedral symmetry ($R\bar{3}c$, LaCoO$_3$, LaNiO$_3$), and the structures are shown in **Figure 1**. Our use of pseudocubic structures provides a good approximation to the average cubic symmetry exhibited by many of these materials at elevated temperatures of T > 500 K,[5] and provides a structurally consistent set of materials to investigate compositional trends in the surface work function. We expect the compositional trends in work function and physics described in this work to also hold under room temperature conditions. However, some quantitative differences in work function should be expected. **Figure 1** also provides examples of the asymmetric (**Figure 1C**, used for $La_{1-x}Sr_xMnO_3$) and symmetric (**Figure 1D** and **Figure 1E**, used for all other materials besides $La_{1-x}Sr_xMnO_3$) surface slabs used for the work function calculations. Additional details on perovskite bulk and surface calculations are provided in **Section 1** of the **Supplementary Information (SI)**.

We performed all of our calculations using Density Functional Theory (DFT) as implemented by the Vienna *ab initio* simulation package (VASP)[72] with a plane wave basis set. We used the hybrid HSE exchange and correlation functional of Heyd, Scuseria and Ernzerhof[18] with Perdew-Burke-Ernzerhof (PBE)-type pseudopotentials[73] utilizing the projector augmented wave (PAW)[74] method for La, Ca, Mg, Ba, Sr, Sc, Ti, V, Cr, Mn, Fe, Co, Ni, Ru and O atoms. The fraction of Hartree-Fock (HF) exchange in the HSE method for each material was obtained from Refs. [20] and [19]. In Refs. [20] and [19], the fraction of HF exchange was fitted to reproduce the experimentally-measured bulk band gap and densities of states from ultraviolet photoemission spectroscopy (UPS) measurements. Thus, the fractions of Hartree-Fock exchange used in our HSE calculations were 0.25 ($LaScO_3$), 0.15 ($LaTiO_3$, $LaCrO_3$, $LaMnO_3$, $LaFeO_3$), 0.125 ($LaVO_3$), 0.05 ($LaCoO_3$) and 0 ($LaNiO_3$). For the band insulators $SrTiO_3$ and $LaAlO_3$, a value of 0.25 is used for the HF exchange fraction.[75, 76] For the remaining materials, the HF exchange values used were the same as the respective transition metal-containing lanthanide perovskite. Therefore, for $SrVO_3$, $SrFeO_3$, $SrCoO_3$, $Ba_{0.5}Sr_{0.5}Co_{0.75}Fe_{0.25}O_3$ (BSCF) and $La_{1-x}Sr_xMnO_3$ (LSM), the HF values used were 0.125, 0.15, 0.05, 0.05 and 0.15, respectively. This method of tuning the amount of HF exchange to reproduce experimental bulk electronic structure properties such as the band gap has recently been shown to provide more accurate Li insertion voltages (a quantity that depends sensitively on the electronic structure near the Fermi level) than the default HF exchange of 0.25 for a wide range of transition metal oxide materials.[77] Lastly, the HF exchanges for $LaRuO_3$ and $SrRuO_3$ have, to our knowledge, not been thoroughly characterized in the same manner as the other perovskites studied here. Therefore, we use an HF value equal to that of LaFeO$_3$ (HF exchange of 0.15) because Fe and Ru are in the same column of the periodic table and thus can be expected to exhibit similar chemistry. Because LaRuO$_3$ and SrRuO$_3$ have not been as widely studied computationally with hybrid functionals, their calculated work functions may show larger errors than the other systems studied here.

The valence electron configurations of the atoms utilized in the calculations were La: $5s^2 5p^6 6s^2 5d^1$, Ca: $3s^2 3p^6 4s^2$, Mg: $2s^2 2p^6 3s^2$, Ba: $5s^2 5p^6 6s^2$, Sr: $3s^2 3p^6 4s^2$, Sc: $3s^2 3p^6 4s^2 3d^1$, Ti: $3s^2 3p^6 4s^2 3d^2$, V: $3p^6 4s^1 3d^4$, Cr: $3p^6 4s^1 3d^5$, Mn: $3p^6 4s^2 3d^5$, Fe: $3s^2 3p^6 4s^1 3d^7$, Co: $4s^1 3d^8$, Ni: $3p^6 4s^2 3d^8$, Al: $3s^2 3p^1$, Ru: $4p^6 5s^1 4d^7$ and O: $2s^2 2p^4$ respectively. The plane wave cutoff energies were, at a minimum, 30% larger than the maximum plane wave energy of the chosen pseudopotentials, and equal to a minimum of 405 eV for all systems. We performed all calculations with spin polarization. The Monkhorst-Pack scheme was used for reciprocal space integration in the Brillouin Zone for bulk perovskite materials.[78] For surface calculations we used a Γ-centered reciprocal space integration scheme instead of Monkhorst-Pack as we use only one k-point, and the electronic minimization was performed simultaneously for all energy bands. A 2x2x2 k-point mesh was used for the 2x2x2 bulk supercells of all LaBO$_3$ materials (40 atoms per cell), with total energy convergence (ionic and electronic degrees of freedom) of 3 meV per formula unit. For surface slab calculations, we reduced the k-point mesh to 1x1x1 and maintained a minimum vacuum distance of 15 Å. We verified that all calculated work functions were well-converged (error of approximately +/- 0.1 eV) with respect to both slab thickness and vacuum region thickness, with the exception of LaAlO$_3$ and LaScO$_3$, which are highly polar materials and with work functions which converge very slowly with slab thickness. Therefore, work function results for LaScO$_3$ and LaAlO$_3$ have a larger error of approximately +/- 0.4 eV, based on GGA calculations of symmetric (001) surface slabs of LaAlO$_3$ between 5 and 17 layers. Lastly, we implemented the dipole correction in VASP to ensure vacuum level convergence, and the dipole correction was calculated only in the axial direction normal to the terminating surface. Additional details regarding the bulk and surface calculations of the perovskite materials considered here can be found in **Section 1** and **Section 2** of the **SI**.

# Supporting Information

Supporting Information is available from the Wiley Online Library or from the author

# Acknowledgements

This work was supported by the US Air Force Office of Scientific Research through grants No. FA9550-08-0052 and No. FA9550-11-0299. Computational support was provided by the Extreme Science and Engineering Discovery Environment (XSEDE), which is supported by National Science Foundation Grant No. OCI-1053575. This research was performed using the compute resources and assistance of the UW-Madison Center For High Throughput Computing (CHTC) in the Department of Computer Sciences.